\documentclass[conference]{IEEEtran}
\IEEEoverridecommandlockouts
\usepackage{cite}
\usepackage{amsmath,amssymb,amsfonts}
\usepackage{graphicx}
\usepackage{textcomp}
\usepackage{xcolor}
\usepackage{hhline}
\usepackage{algorithm}
\usepackage{algpseudocode}
\usepackage{bm}

\def\BibTeX{{\rm B\kern-.05em{\sc i\kern-.025em b}\kern-.08em
    T\kern-.1667em\lower.7ex\hbox{E}\kern-.125emX}}

\IEEEpubid{\makebox[\columnwidth]{978-1-5386-7848-0/18/\$31.00 ©2018 IEEE \hfill} \hspace{\columnsep}\makebox[\columnwidth]{ }}

\newcommand\blfootnote[1]{%
	\begingroup
	\renewcommand\thefootnote{}\footnote{#1}%
	\addtocounter{footnote}{-1}%
	\endgroup
}

\begin{document}

\title{Early Identification of Pathogenic Social Media Accounts\\
}

\author{\IEEEauthorblockN{Hamidreza Alvari, Elham Shaabani, Paulo Shakarian}
\textit{Arizona State University}\\
Tempe, USA \\
\{halvari, eshaaban, shak\}@asu.edu}

\maketitle

\begin{abstract}
Pathogenic Social Media (PSM) accounts such as terrorist supporters exploit large communities of supporters for conducting attacks on social media. Early detection of these accounts is crucial as they are high likely to be \textit{key} users in making a harmful message ``viral''. In this paper, we make the first attempt on utilizing causal inference to identify PSMs within a short time frame around their activity. We propose a time-decay causality metric and incorporate it into a causal community detection-based algorithm. The proposed algorithm is applied to groups of accounts sharing similar causality features and is followed by a classification algorithm to classify accounts as PSM or not. Unlike existing techniques that take significant time to collect information such as network, cascade path, or content, our scheme relies solely on action log of users. Results on a real-world dataset from Twitter demonstrate \textit{effectiveness} and \textit{efficiency} of our approach. We achieved precision of 0.84 for detecting PSMs only based on their first 10 days of activity; the misclassified accounts were then detected 10 days later. 
\end{abstract}


\section{Introduction}
\blfootnote{U.S. Provisional Patent 62/628,196. Contact shak@asu.edu for licensing information.}The unregulated nature and rapid growth of the Web have raised numerous challenges, including hate speech~\cite{badjatiya2017deep}, human trafficking~\cite{alvari2017semi} and disinformation spread~\cite{Causal2017} which ultimately pose threats to users privacy~\cite{beigi2018securing,beigi2018privacy}. Take disinformation spread as an example where ``Pathogenic Social Media'' (PSM) accounts (e.g., terrorist supporters, or fake news writers)~\cite{Causal2017} seek to promote or degrade certain ideas by utilizing large online communities of supporters to reach their goals. Identifying PSMs has applications including countering terrorism~\cite{khader2016combating,DBLP:journals/corr/KlausenMZ16}, fake news detection~\cite{Gupta2014,6805772} and water armies detection~\cite{DBLP:journals/corr/abs-1111-4297}.

Early detection of PSMs in social media is crucial as they are likely to be \textit{key} users to malicious campaigns~\cite{DBLP:journals/corr/VarolFMF17}. This is a challenging task for three reasons. First, these platforms are primarily based on reports they receive from their own users
\footnote{https://bit.ly/2Dq5i4M} to manually shut down PSMs which is not a timely approach. Despite efforts to suspend these accounts, many of them simply return to social media with different accounts. Second, the available data is often imbalanced and social network structure, which is at the core of many techniques~\cite{conf/icwsm/WengMA14,Kempe2003,beigi2018similar,Zhang:2013:IIN:2444040.2444143}, is not readily available. Third, PSMs often seek to utilize and cultivate large number of online communities of passive supporters to spread as much harmful information as they can. 




\noindent\textbf{Present Work.} Causal inference is tailored to identify PSMs since they are key users in making a harmful message ``viral''-- where ``viral'' is defined as an order-of-magnitude increase. We propose \textit{time-decay} causal metrics to distinguish PSMs from normal users within a \textit{short} time around their activity. Our metrics alone can achieve high classification performance in identification of PSMs soon after they perform actions. 
Next, we pose the following research question: \textit{Are causality scores of users within a community higher than those across different communities?} We propose a \underline{c}ausal \underline{c}ommunity \underline{d}etection-based \underline{c}lassification method (\textsc{C$^2$dc}), that takes causality of users and the community structure of their action log. 

\noindent\textbf{Contributions.} We make the following major contributions:
\begin{itemize}	
	\item We enrich the causal inference framework of~\cite{DBLP:journals/corr/abs-1205-2634} and present \textit{time-decay} extensions of the causal metrics in~\cite{Causal2017} for early identification of PSMs. 
	
	\item We investigate the role of community structure in early detection of PSMs, by demonstrating that users within a community establish stronger causal relationships compared to the rest. 
	
	\item We conduct a suit of experiments on a dataset from Twitter. Our metrics reached F1-score of 0.6 in identifying PSMs, half way their activity, and identified 71\% of PSMs based on first 10 days of their activity, via supervised settings. The community detection approach achieved precision of 0.84 based on first 10 days of users activity; the misclassified accounts were identified based on their activity of 10 more days. 
\end{itemize}


\section{Background}

\subsection{Technical Preliminaries}
Following the convention of~\cite{Goyal:2010}, we assume an \textit{action log} $\mathbf{A}$ of the form \textit{Actions(User,Action,Time)}, which contains tuples $(u,a_u,t_u)$ indicating that user $u$ has performed action $a_u$ at time $t_u$. For ease of exposition, we slightly abuse the notation and use the tuple $(u,m,t)$ to indicate that user $u$ has posted (tweeted/retweeted) message $m$ at time $t$. For a given message $m$ we define a \textit{cascade} of actions as $\mathbf{A}_m= \{(u,m',t)\in\mathbf{A}|m'=m\}$.

User $u$ is said to be an $m$-participant if there exists $t_u$ such that $(u,m,t_u)\in\mathbf{A}$. For users who have adopted a message in the early stage of its life span, we define \textit{key users} as follows.

\textbf{Definition 1 (Key Users).} \textit{Given message $m$, $m$-participant $u$ and cascade $\mathbf{A}_m$, we say user $u$ is a key user iff user $u$ precedes at least $\phi$ fraction of other $m$-participants where $\phi\in(0,1)$. In other words,} $|\mathbf{A}_m|\times\phi \leq |\{j|\exists t': (j,m,t')\in \mathbf{A}\wedge t < t'\}|$, \textit{where $|.|$ is the cardinality of a set}. 

Next, we define viral messages as follows.

\textbf{Definition 2 (Viral Messages).} \textit{Given a threshold $\theta$, we say a message $m\in \mathbf{M}$ is viral iff $|\mathbf{A}_m| \geq \theta$. We denote a set of all viral messages by $\mathbf{M}_{vir}$.}

The prior probability of a message going viral is $\rho = |\mathbf{M}_{vir}|/|\mathbf{M}|$. The probability of a message going viral given key user $u$ has participated in, is computed as follows:

\begin{equation}\small
	\rho_u = \frac{|\{m|m\in \mathbf{M}_{vir}\wedge \textit{u is a key user}\}|}{|\{m|m\in \mathbf{M} \wedge \textit{u is a key user}\}|}
\end{equation}

The probability that key users $i$ and $j$ tweet/retweet message $m$ chronologically and make it viral, is computed as:

\begin{equation}\small
	p_{i,j}=\frac{|\{m\in \mathbf{M}_{vir}|\exists t, t': t<t' \wedge (i,m,t), (j,m,t') \in \mathbf{A}\}|}{|\{m\in \mathbf{M}|\exists t, t': t<t' \wedge (i,m,t), (j,m,t') \in \mathbf{A}\}|}
\end{equation}

To examine how causal user $i$ was in helping a message $m$ going viral, we shall explore what will happen if we exclude user $i$ from $m$. We define the probability that \textit{only} key user $j$ has made a message $m$ viral, i.e. user $i$ has not posted $m$ or does not precede $j$ as: 

\begin{equation}\small
	p_{\neg i,j}=
	\frac{|\{m\in \mathbf{M}_{vir}|\exists t': (j,m,t') \in \mathbf{A} \wedge \nexists t:t< t', (i,m,t) \in \mathbf{A}\}|}{|\{m\in \mathbf{M}|\exists t': (j,m,t') \in \mathbf{A} \wedge \nexists t:t< t', (i,m,t) \in \mathbf{A}\}|}
\end{equation}

Next, we adopt the notion of Prima Facie causes~\cite{suppes1970}:

\textbf{Definition 3 (Prima Facie Causal Users).} \textit{A user $u$ is said to be Prima Facie causal user for cascade $\mathbf{A}_m$ iff: (1) user $u$ is a key user of $m$, (2) $m \in \mathbf{M}_{vir}$, and (3) $\rho_u>\rho$.
}

We borrow the concept of \textit{related users} from a rule-based system~\cite{DBLP:journals/corr/StantonTJVCS15} which was an extension to the causal inference framework in~\cite{DBLP:journals/corr/abs-1205-2634}. We say users $i$ and $j$ are $m$-related if (1) both are Prima Facie causal for $m$, and (2) $i$ precedes $j$. We then define a set of user $i$'s related users as $\mathbf{R}(i) = \{j|j\ne i \textit{ and } i, j \textit{ are \textit{m}-related}\}$.



\begin{figure*}[t]\center
	\includegraphics[width=0.2\textwidth]{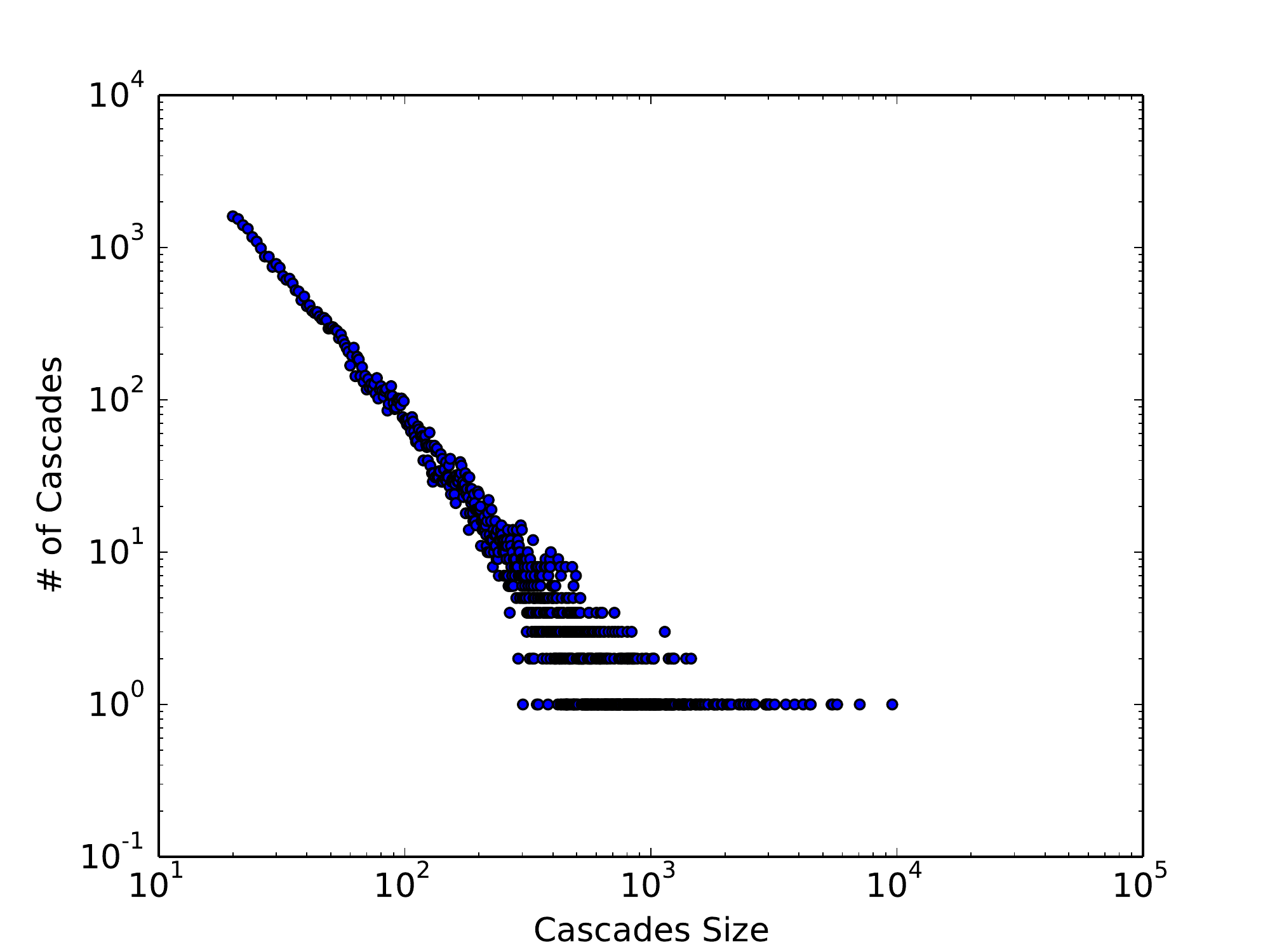}
	\includegraphics[width=0.2\textwidth]{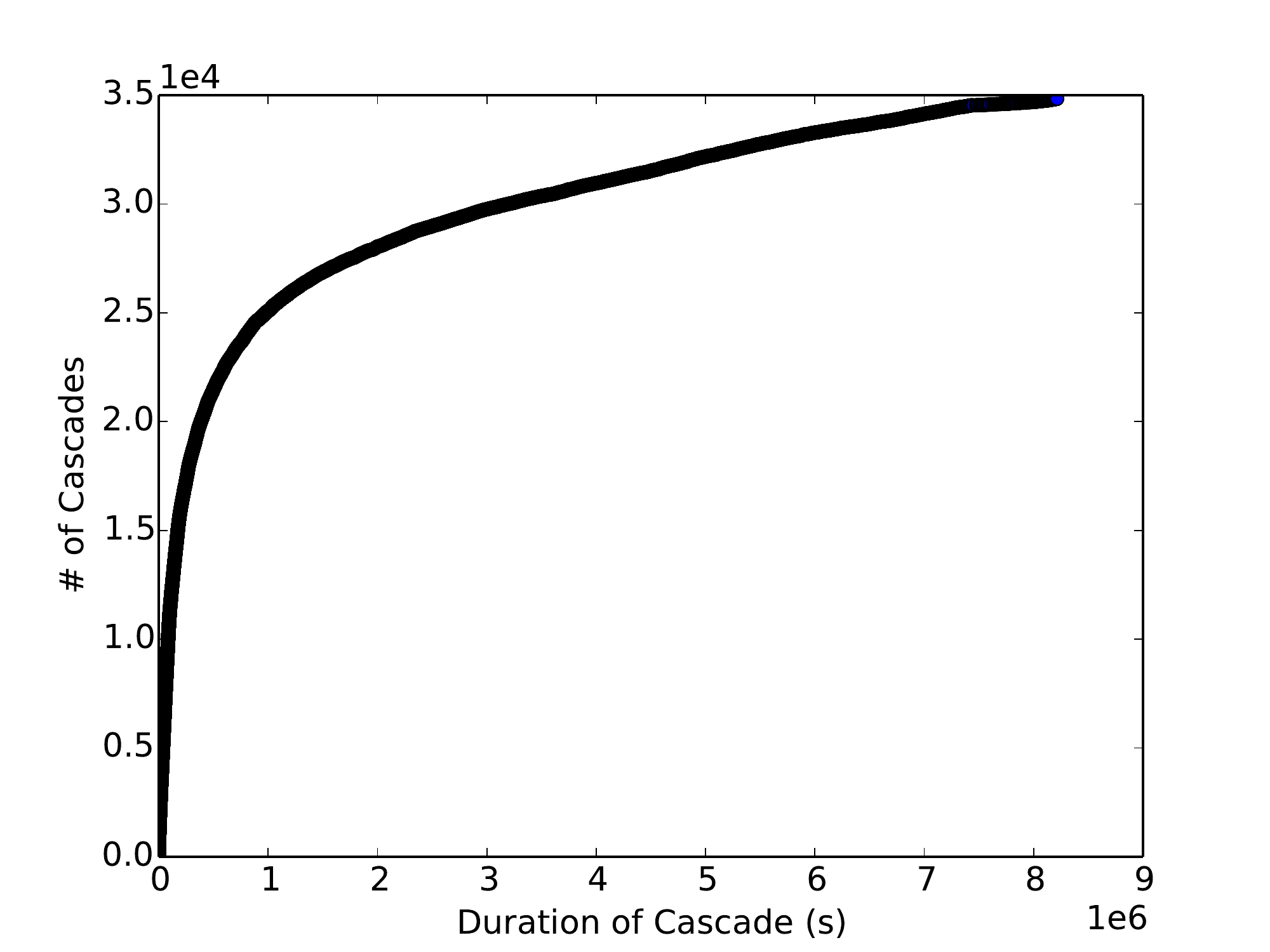}	
	\includegraphics[width=0.2\textwidth]{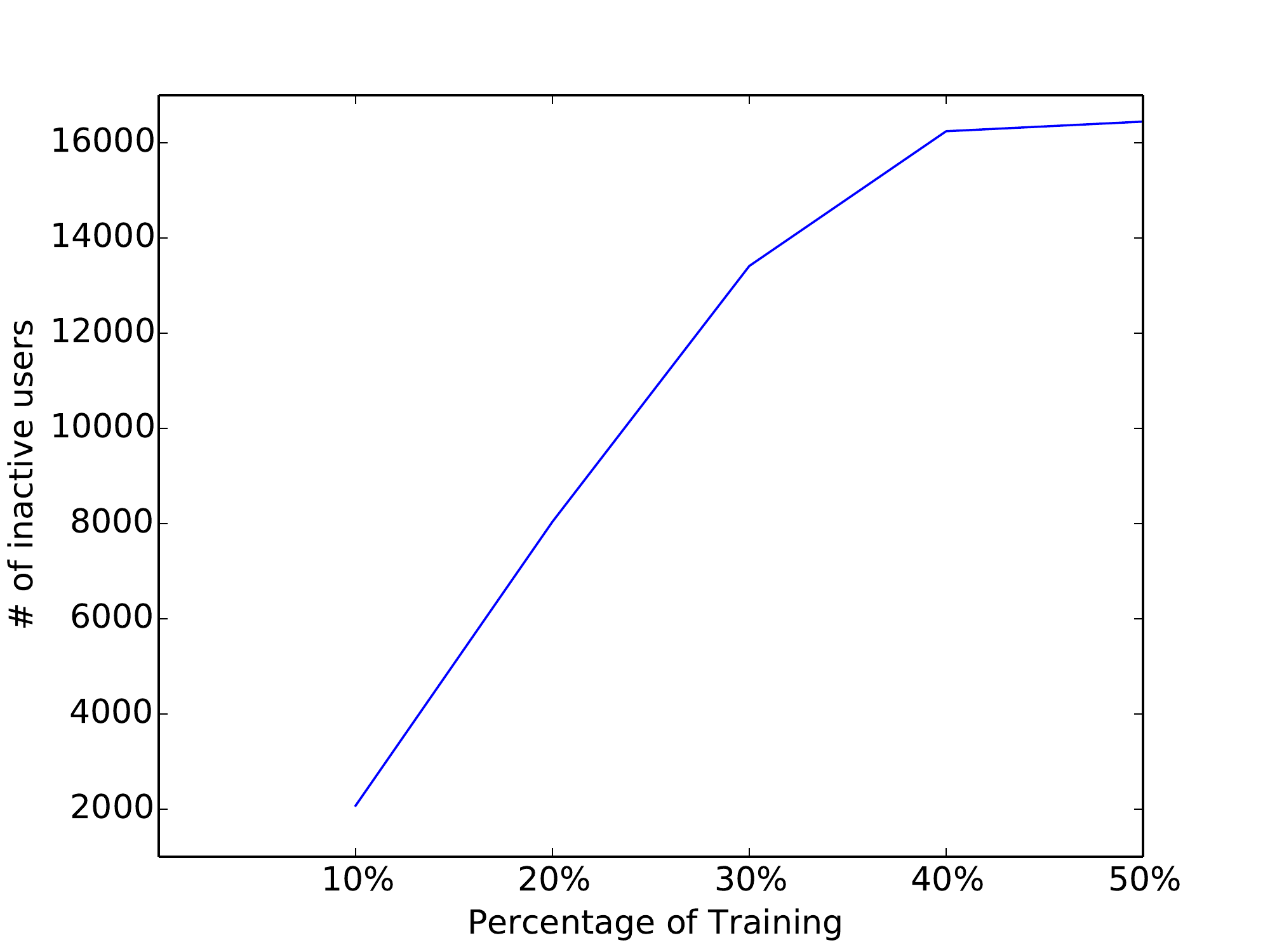}
	\includegraphics[width=0.2\textwidth]{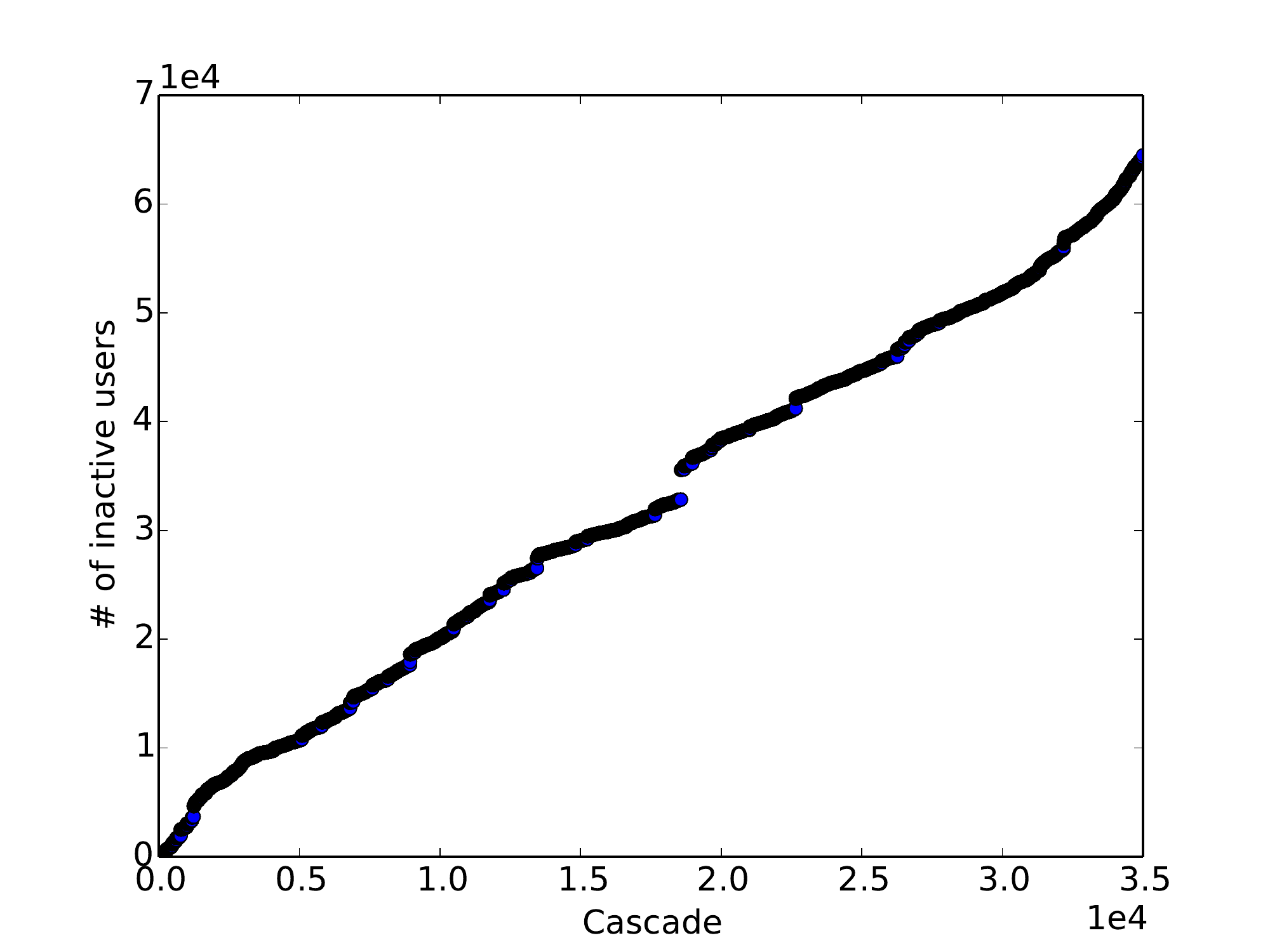}
	\caption{Left to right: Log-log distribution of cascades vs. cascade size. Cumulative distribution of duration of cascades. Number of inactive users in different subsets of the training set. Total inactive users in each cascade.}
	\label{fig:data_analysis}
\end{figure*}


\subsection{Dataset}
We collect a dataset (Table~\ref{tb:st}) of 53M ISIS related tweets/retweets in Arabic, from Feb 22, 2016 to May 27, 2016. 
The dataset has different fields including user ID, retweet ID, hashtags, content, posting time. 
The tweets were collected using 290 different hashtags such as \textsf{\#Terrorism} and \textsf{\#StateOfTheIslamicCaliphate}. We use a subset of this dataset which contains 35K cascades of different sizes and durations. There are $\sim$2.8M tweets/retweets associated with the cascades. After pre-processing and removing duplicate users from cascades, cascades sizes (i.e. number of associated postings) vary between 20 to 9,571 and take from 10 seconds to 95 days to finish. The log-log distribution of cascades vs. cascade size and the cumulative distribution of duration of cascades are depicted in Figure~\ref{fig:data_analysis}. 

Based on the content of tweets in our dataset, PSMs are terrorism-supporting accounts who have participated in viral cascades. We chose to use $\theta=100$ and take $\sim$6K viral cascades with at least 100 tweets/retweets. We demonstrate number of PSMs that have been suspended by the Twitter over time and total number of suspended users in each cascade, in Figure~\ref{fig:data_analysis}. We experiment the effectiveness of our proposed approach on subsets of the training set with different sizes. Note we use no more than 50\% of original dataset to ensure our approach is able to identify PSMs early enough. The dataset does not have any underlying network. We only focus on the non-textual information in the form of an \textit{action log}. We set $\phi=0.5$ to select \textit{key users} and after the data collection, we check through Twitter API whether they have been suspended (PSM) or they are active (normal)~\cite{thomas2011suspended}. According to Table~\ref{tb:st}, 11\% of the users in our dataset are PSM and others are normal.

\begin{table}[t]\small
	\centering
	\caption{Description of the dataset.}
	\begin{tabular}{|l|c|c|}
		\cline{1-3}
		\textbf{Name}          & \multicolumn{2}{c|}{\textbf{Value}}\\
		\hhline{===}
		\# of Cascades      & \multicolumn{2}{c|}{35K}  \\ \cline{1-3}
		\# of Viral Cascades & \multicolumn{2}{c|}{6,602} \\ \cline{1-3}
		\# of Tweets/Retweets & \multicolumn{2}{c|}{2,808,878} \\ \cline{1-3}
		\# of Users  & Suspended & Active \\ \cline{2-3}
		& 64,484
		& 536,609\\
		\cline{1-3}
	\end{tabular}
	\label{tb:st}
\end{table}

\subsection{Causal Measures}
Causal inference framework was first introduced in~\cite{DBLP:journals/corr/abs-1205-2634}. Later,~\cite{Causal2017} adopted the framework and extended it to suite the problem of identifying PSMs. They extend the Kleinberg-Mishra causality ($\epsilon_{K\&M}$) to a series of causal metrics. To recap, we briefly explain them in the following discussion. Before going any further, $\epsilon_{K\&M}$ is computed as follows:

\begin{equation}\small
	\epsilon_{K\&M}(i)=\frac{ \sum_{j\in \mathbf{R}(i)}(p_{i,j}-p_{\neg i,j}) }{|\mathbf{R}(i)|}
\end{equation}  

This metric measures how causal user $i$ is, by taking the average of $p_{i,j}-p_{\neg i,j}$ over $\mathbf{R}(i)$. The intuition here is user $i$ is more likely to be cause of message $m$ to become viral than user $j$, if $p_{i,j}-p_{\neg i,j} > 0$. The work of~\cite{Causal2017} devised a suit of the variants, namely relative likelihood causality ($\epsilon_{rel}$), neighborhood-based causality ($\epsilon_{nb}$) and its weighted version ($\epsilon_{wnb}$). Note that none of these metrics were originally introduced for \textit{early} identification of PSMs. Therefore, we shall make slight modifications to their notations to adjust our temporal formulations, using calligraphic uppercase letters. We define $\mathcal{E}_{K\&M}$ over a given time interval $I$ as follows: 

\begin{equation}\small
	\mathcal{E}^I_{K\&M}(i)=\frac{ \sum_{j\in \mathcal{R}(i)}(\mathcal{P}_{i,j}-\mathcal{P}_{\neg i,j}) }{|\mathcal{R}(i)|}
\end{equation}

\noindent where $\mathcal{R}(i)$, $\mathcal{P}_{i,j}$, and $\mathcal{P}_{\neg i,j}$ are now defined over $I$. Authors in~\cite{Causal2017} mention that this metric cannot spot all PSMs. They define another metric, relative likelihood causality $\mathcal{E}_{rel}$, which works by assessing relative difference between $\mathcal{P}_{i,j}$, and $\mathcal{P}_{\neg i,j}$. We use its temporal version over $I$, $\mathcal{E}^I_{rel}(i)=\frac{ \mathcal{S}(i,j) }{|\mathcal{R}(i)|}$.

\noindent where $\mathcal{S}(i,j)$ is defined as follows and $\alpha$ is infinitesimal:

\begin{equation}\small
	\mathcal{S}(i,j)=\begin{cases}
		\frac{\mathcal{P}_{i,j}}{\mathcal{P}_{\neg i,j} + \alpha}-1, & \mathcal{P}_{i,j} > \mathcal{P}_{\neg i,j}\\
		1- \frac{\mathcal{P}_{\neg i,j}}{\mathcal{P}_{i,j}}, & \mathcal{P}_{i,j} \leq \mathcal{P}_{\neg i,j}
	\end{cases}
\end{equation}

Two other neighborhood-based metrics were also defined in~\cite{Causal2017}, whose temporal variants are computed over $I$ as $\mathcal{E}^I_{nb}(j)=\frac{ \sum_{i\in \mathcal{Q}(j)}\mathcal{E}^I_{K\&M}(i) }{|\mathcal{Q}(j)|}$, where $\mathcal{Q}(j)=\{i|j\in \mathcal{R}(i)\}$ is the set of all users that user $j$ belongs to their related users sets.
Similarly, the second metric is a weighted version of the above metric and is called weighted neighborhood-based causality and is calculated as $\mathcal{E}^I_{wnb}(j)=\frac{ \sum_{i\in \mathcal{Q}(j)}w_i\times\mathcal{E}^I_{K\&M}(i) }{\sum_{i\in \mathcal{Q}(j)}w_i}$. This is to capture different impacts that users in $Q(j)$ have on user $j$.
We apply a threshold-based selection approach that selects PSMs from normal users, based on a given threshold. Following~\cite{Causal2017}, we use a threshold of 0.7 for all metrics except $\mathcal{E}^I_{rel}$ for which we used a threshold of 7 (Table~\ref{tb:f1-causal}). 


\begin{table}\small
	\centering
	\caption{F1-score results for PSM accounts using each causal metric in~\cite{Causal2017}.}
	\begin{tabular}{|c|c|c|c|c|c|} 
		\cline{1-6}
		\textbf{Metric} & \multicolumn{5}{c|}{\textbf{F1-score}} \\
		\hhline{======}
		& 10\% & 20\% & 30\% & 40\% & 50\% \\ \cline{1-6}
		$\mathcal{E}^I_{K\&M}$ & 0.41 & 0.42 & 0.45 & 0.46 & 0.49 \\ \hline
		$\mathcal{E}^I_{rel}$ & 0.3 & 0.31 & 0.33 & 0.35 & 0.37 \\ \hline
		$\mathcal{E}^I_{nb}$ & 0.49 & 0.51 & 0.52 & 0.54 & 0.55 \\ \hline
		$\mathcal{E}^I_{wnb}$ & \textbf{0.51} & \textbf{0.52} & \textbf{0.55} & \textbf{0.56} & \textbf{0.59} \\ \hline
	\end{tabular}
	\label{tb:f1-causal}
\end{table}

\begin{figure}\center \small
	\includegraphics[width=0.3\textwidth]{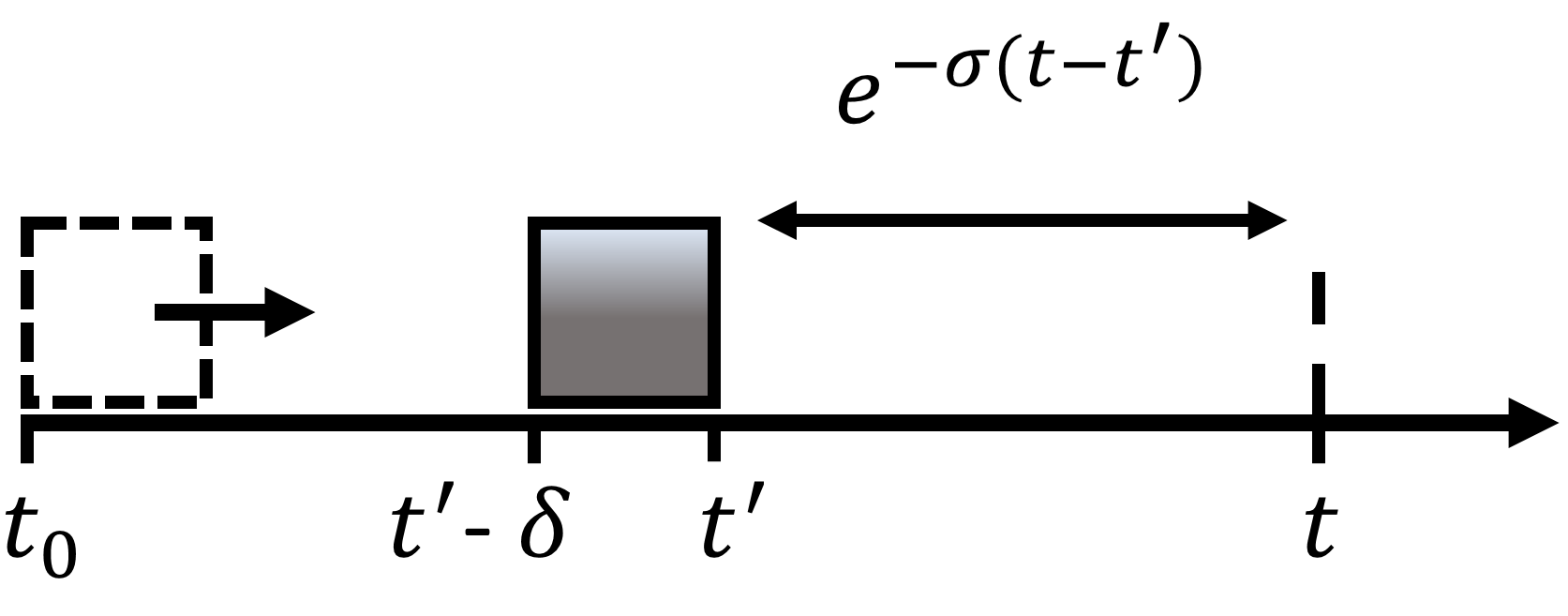}
	\caption{An illustration of how decay-based causality works. To compute $\xi_{k}^I(i)$ over $I=[t_0,t]$, we use a sliding window $\Delta=[t'-\delta, t']$ and take the average between the resultant causality scores $e^{-\sigma (t-t')}\times\mathcal{E}^{\Delta}_{k}(i)$.}
	\label{fig:decay_ex}
\end{figure}

\section{The proposed Framework}
\subsection{Leveraging Temporal Aspects of Causality}
Previous causal metrics do not take into account time-decay effect. They assume a steady trend for computing causality scores. This is an unrealistic assumption, as causality of users may change over time. We introduce a generic decay-based metric. Our metric assigns different weights to different time points of a given time interval, inversely proportional to their distance from $t$ (i.e., smaller distance is associated with higher weight). Specifically, it performs the following: it (1) breaks down the given time interval into shorter time periods using a sliding time window, (2) deploys an exponential decay function of the form $f(x)=e^{-\alpha x}$ to account for the time-decay effect, and (3) takes average of the causality values computed over each sliding time window. Formally, $\xi^I_{k}$ is defined as follows, where $k\in\{K\&M,rel,nb,wnb\}$:   

\begin{equation}\small
	\xi_{k}^I(i)=\frac{1}{|\mathcal{T}|}\sum_{t'\in \mathcal{T}}e^{-\sigma (t-t')}\times\mathcal{E}^{\Delta}_{k}(i)
\end{equation}

\noindent where $\sigma$ is a scaling parameter of the exponential decay function, $\mathcal{T}=\{t'|t'=t_0+j\times \delta, j\in\mathbb{N} \wedge t'\leq t-\delta\}$ is a sequence of sliding-time windows, and $\delta$ is a small fixed amount of time, which is used as the length of each sliding-time window $\Delta=[t'-\delta, t']$ (Figure~\ref{fig:decay_ex}).     
To apply the threshold-based approach, we once again use a threshold of 0.7 for all metrics except $\xi^I_{rel}$ for which we used a threshold of 7 (Table~\ref{tb:f1-decay}). 


\begin{table*}\small
	\centering
	\caption{F1-score results for PSM accounts using each decay-based metric with and without communities.}
	\begin{tabular}{|c|c|c|c|c|c|} 
		\cline{1-6}
		\textbf{Metric} & \multicolumn{5}{c|}{\textbf{F1-score (without/with communities)}} \\
		\hhline{======}
		& 10\% & 20\% & 30\% & 40\% & 50\% \\ \cline{1-6}
		$\xi^I_{K\&M}$ & 0.44/0.49 & 0.46/0.51 & 0.47/0.52 & 0.5/0.54 & 0.53/0.57 \\ \hline
		$\xi^I_{rel}$ & 0.36/0.4 & 0.38/0.43 & 0.39/0.46 & 0.41/0.49 & 0.42/0.5 \\ \hline
		$\xi^I_{nb}$ & 0.52/0.56 & 0.53/0.57 & 0.54/0.58 & 0.56/0.6 & 0.59/0.61 \\ \hline
		$\xi^I_{wnb}$ & \textbf{0.54/0.57} & \textbf{0.55/0.58} & \textbf{0.57/0.6} & \textbf{0.58/0.62} & \textbf{0.6/0.63} \\ \hline
	\end{tabular}
	\label{tb:f1-decay}
\end{table*}


\textbf{Early Detection of PSMs.} \textit{Given action log $\mathbf{A}$, and user $u$ where $\exists t$ \textit{s.t.} $(u,m,t)\in\mathbf{A}$, our goal is to determine if $u$'s account shall be suspended given its causality vector $\mathbf{x}_u\in \mathbb{R}^d$ (here, $d=4$) computed using any of the causality metrics over $[t-\delta,t+\delta]$.    
}


\subsection{Leveraging Community Structure Aspects of Causality}

\begin{figure*}[t]\center \small
	\includegraphics[width=0.21\textwidth]{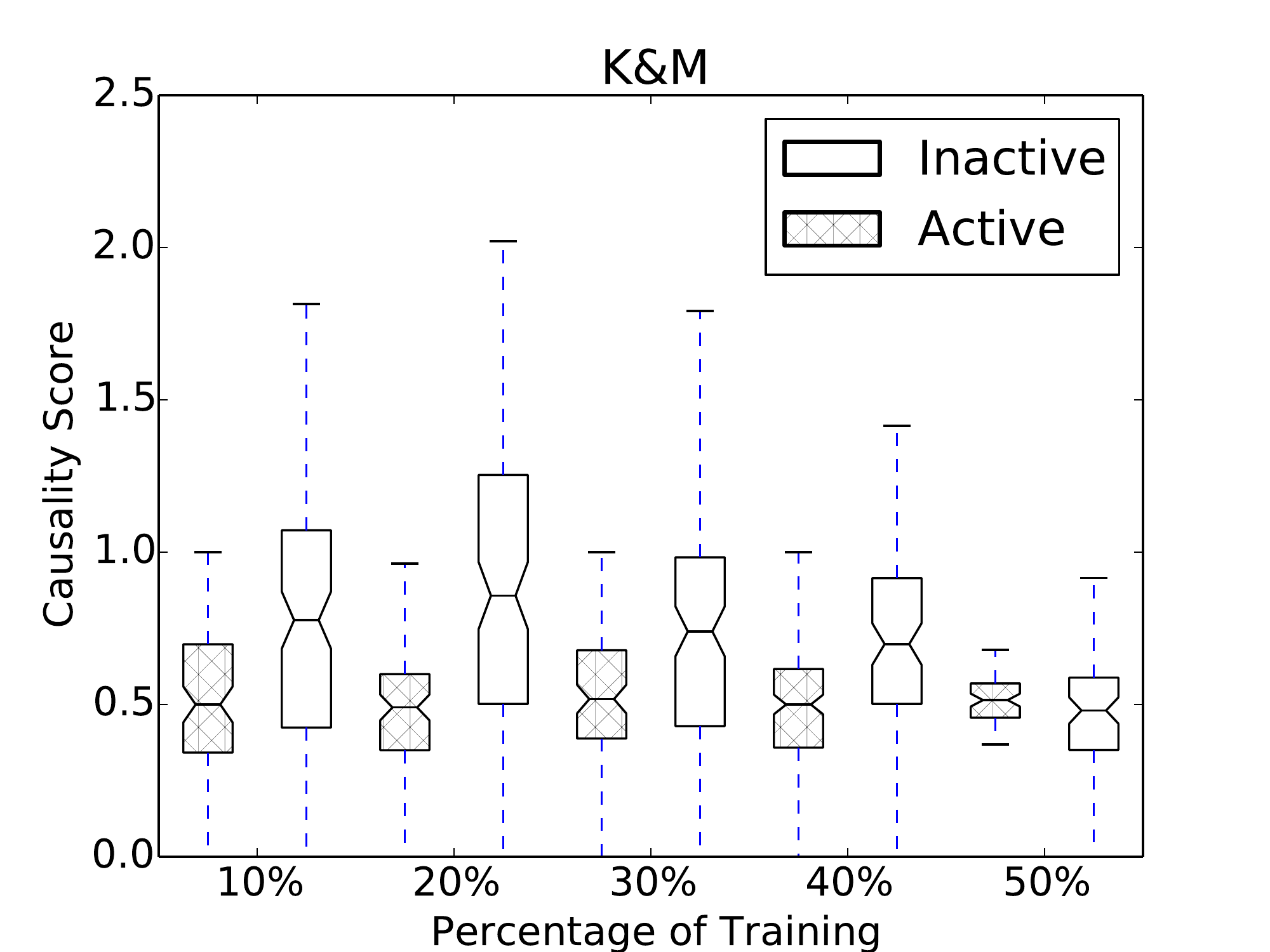}
	\includegraphics[width=0.21\textwidth]{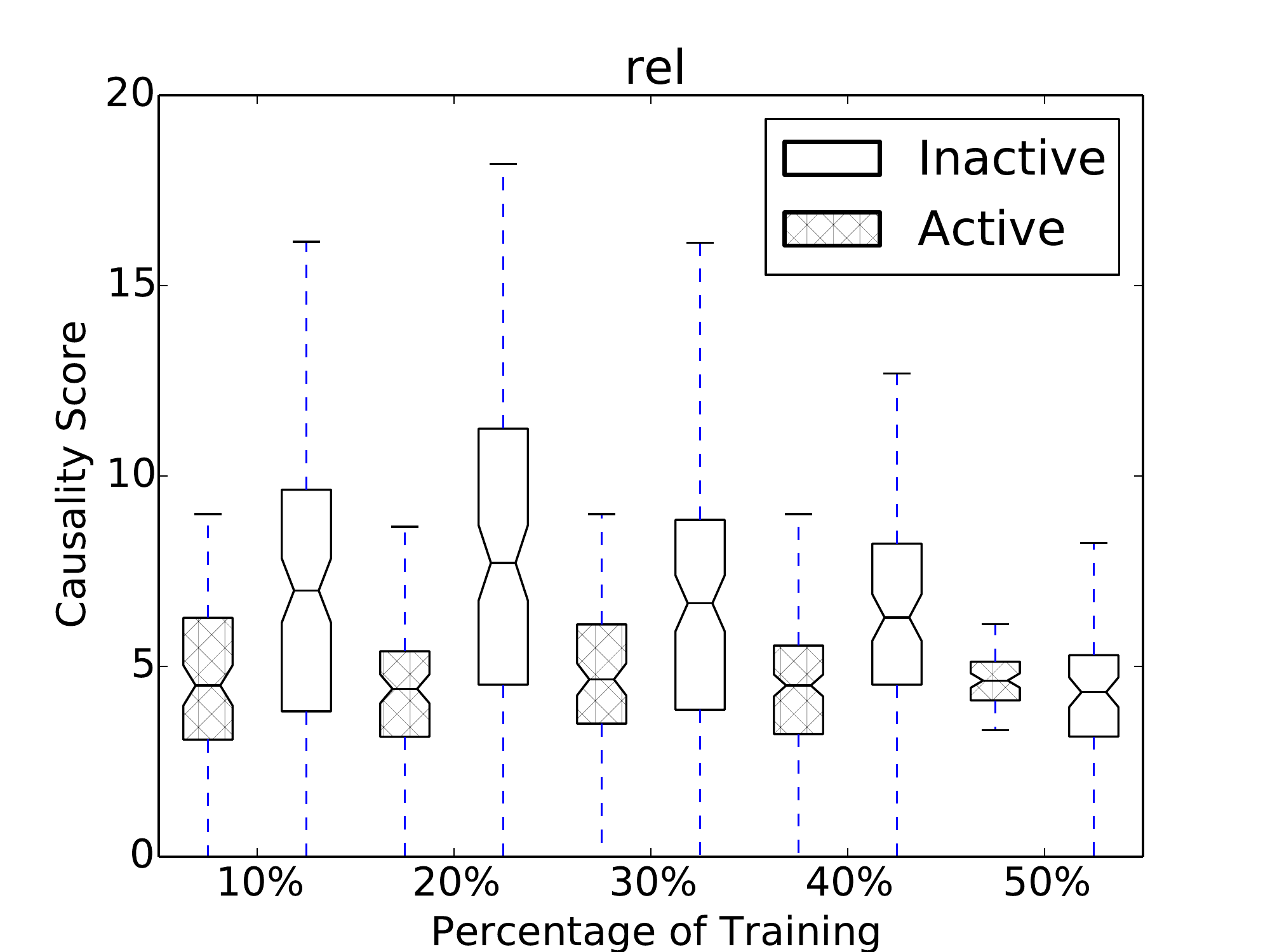}
	\includegraphics[width=0.21\textwidth]{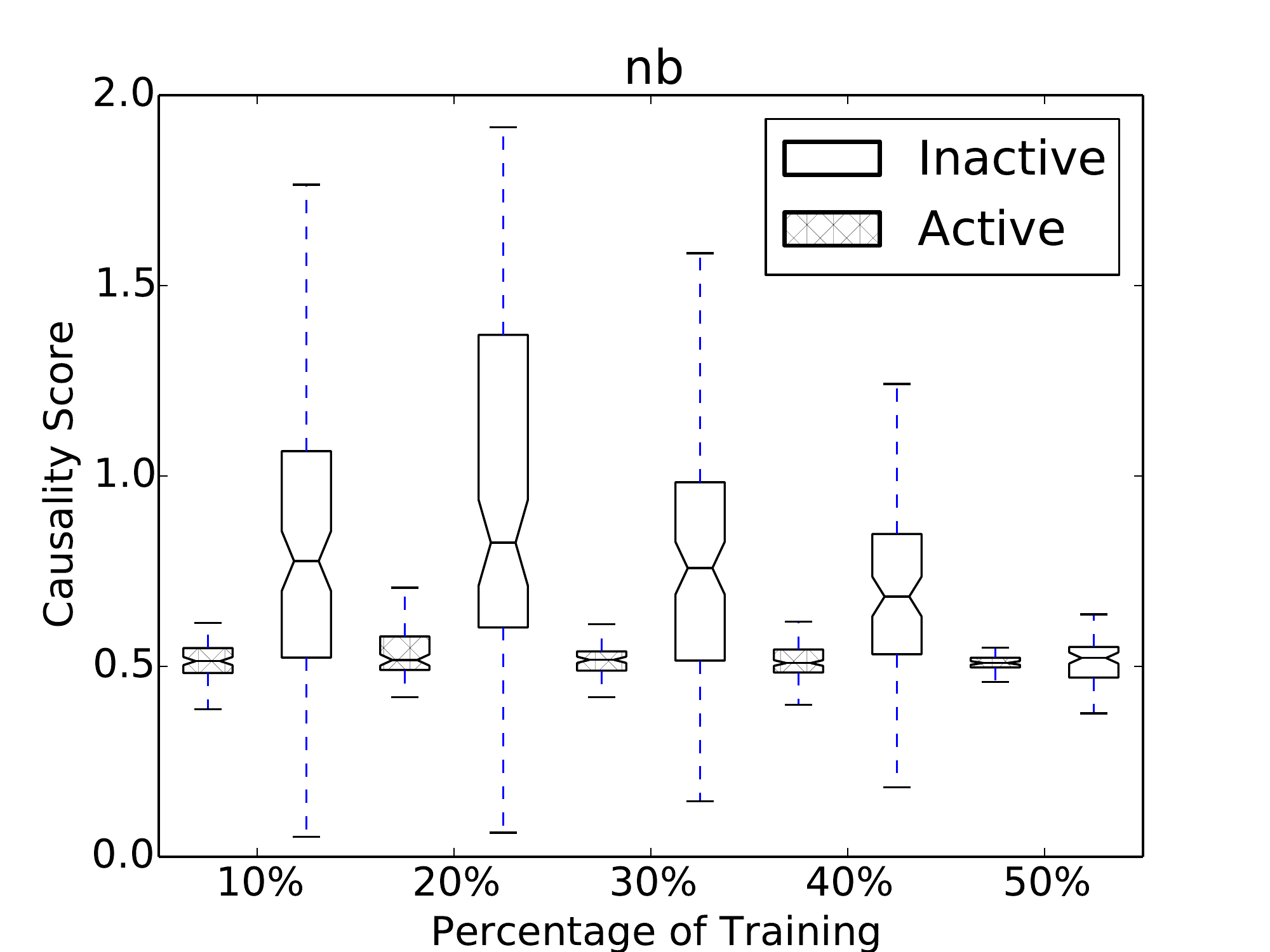}
	\includegraphics[width=0.21\textwidth]{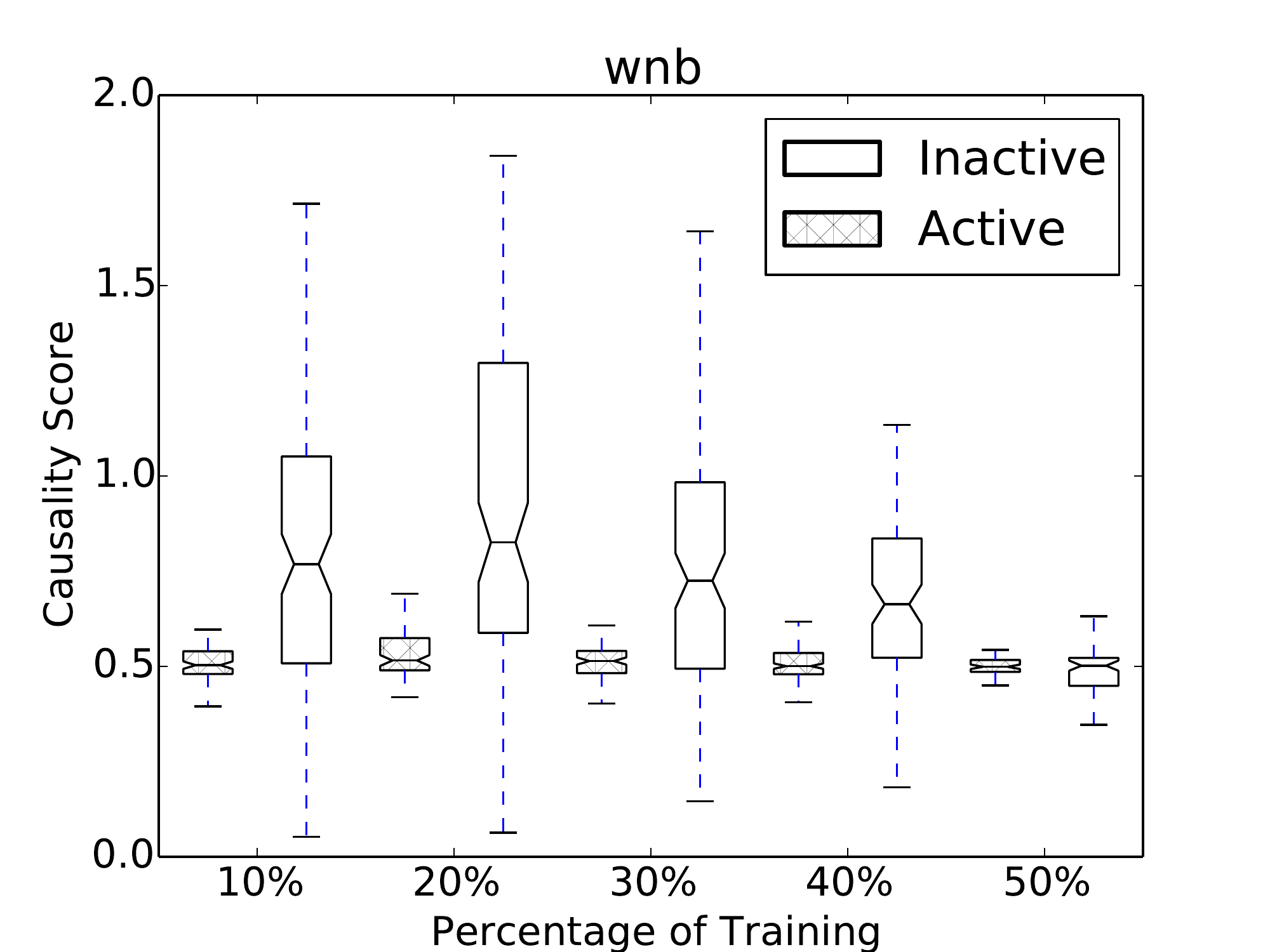}
	\caption{Left to right: distributions of active and inactive users using communities and $\xi^I_{k}$ when $k\in\{K\&M, rel, nb, wnb\}$.}
	\label{fig:dist-community}
\end{figure*}


To answer the research question posed earlier, since network structure is not available, we need to build a graph $\mathbf{G}=(\mathbf{V},\mathbf{E})$ from $\mathbf{A}$ by connecting any pairs of users who have posted same message \textit{chronologically}. In this graph, $\mathbf{V}$ is a set of vertices (i.e. users) and $\mathbf{E}$ is a set of directed edges between users. For the sake of simplicity and without loss of generality, we make the edges of this graph undirected. Next, we leverage the \textsc{Louvain} algorithm~\cite{1742-5468-2008-10-P10008} to find the partitions $\mathbf{C}=\{C_1, C_2, ..., C_k\}$ of $k$ communities over $\mathbf{G}$. Among a myriad of the community detection algorithms~\cite{alvari2016identifying,lancichinetti2011finding,alvari2011detecting}, we chose \textsc{Louvain} due to its fast runtime and scalability-- we leave examining other community detection algorithms to future work. Next, we perform the two-sample $t$-test $H_0:v_a\geq v_b,~H_1:v_a<v_b$. The null hypothesis is: \textit{users in a given community establish weak causal relations with each other as opposed to the other users in other communities}. We construct two vectors $v_a$ and $v_b$ as follows. We create $v_a$ by computing Euclidean distances between causality vectors $(\mathbf{x}_i,\mathbf{x}_j)$ corresponding to each pair of users $(u_i,u_j)$ who are from same community $C_l\in \mathbf{C}$. Therefore, $v_a$ contains exactly $\frac{1}{2}\sum_{l=1}^{|\mathbf{C}|}{|C_l|.(|C_l|-1)}$ elements. We construct $v_b$ of size $\sum_{l=1}^{|\mathbf{C}|}{|C_l|}$ by computing Euclidean distance between each user $u_i$ in community $C_l\in \mathbf{C}$, and a random user $u_k$ chosen from the rest of the communities, i.e., $\mathbf{C}\setminus C_l$. The null hypothesis is rejected at significance level $\alpha=0.01$ with the $p$-value of 4.945e-17. We conclude that users in same communities are more likely to establish stronger causal relationships with each other than the rest of the communities. The answer to the question is thus positive. For brevity, we only reported results for 10\% of the training set, while making similar arguments for other percentages is straightforward. Figure~\ref{fig:dist-community} shows box plots of the distributions of users using the decay-based metrics and the communities and same set of thresholds as before. We observe a clear distinction between active/suspended accounts, using the community structure. Results in Table~\ref{tb:f1-decay} show improvements over previous ones.

\begin{algorithm}\small
	\caption{\textbf{Causal Community Detection-Based Classification Algorithm (\textsc{C$^2$dc})}}
	\begin{algorithmic}[1]
		\Require Training samples $\{\mathbf{x}_1,...,\mathbf{x}_N\}$ , tests $\{\mathbf{x}'_1,...,\mathbf{x}'_n\}$, $\mathbf{G}$, $k$
		\Ensure Predicted labels $\{y'_1,...,y'_n\}$
		\State $\mathbf{C}\gets$ \textsc{Louvain}($\mathbf{G})$
		\For{each $\mathbf{x}'_i$}		
		\State $C_l\gets C'\in \mathbf{C}$ s.t. $\mathbf{x}'_i \in C'$
		\State $\mathbf{D}\gets\{\}$
		\For{each $\mathbf{x}_j\in C_l$}
		\State $d_{ij}\gets||\mathbf{x}'_i-\mathbf{x}_j||_2$
		\State $\mathbf{D}\gets\mathbf{D}\cup \{d_{ij}\}$
		\EndFor
		\State $\mathbf{K}\gets \textsc{Knn}(\mathbf{D}$, $k$)
		\State \text{$y'_i\gets$ \textsc{Dominant-Label}($\mathbf{K})$}
		\EndFor
		
	\end{algorithmic}\label{alg:alg1}
\end{algorithm}


First step of the proposed algorithm (Algorithm~\ref{alg:alg1}) involves finding the communities. In the second step, each unlabeled user is classified based on the available labels of her nearby peers in the same community. We use the \textsc{K-Nearest Neighbors (Knn)} algorithm to compute her $k$ nearest neighbors in the same community, based on Euclidean distances between their causality vectors. We label her based on the majority class of her $k$ nearest neighbors in the community. The merit of using community structure over merely using \textsc{Knn} is, communities can give finer-grained and more accurate sets of neighbors sharing similar causality scores.

\section{Experiments}

We use different subsets of size $x$\% of the entire time-line (from Feb 22, 2016 to May 27, 2016) of the action log $\mathbf{A}$, by varying $x$ as $\{10,20,30,40,50\}$. For each subset and user $i$ in the subset, we compute feature vector $\mathbf{x}_i\in\mathbb{R}^4$ of the corresponding causality scores. The feature vectors are then fed into supervised classifiers and the community detection-based algorithm. For the sake of fair comparison, we perform this for both causal and decay-based metrics. For both metrics, we empirically found that $\rho=0.1$ and $\alpha=0.001$ work well. For the decay-based causality metric we shall also assume a sliding window of size of 5 days (i.e. $\delta = 5$) and set $\sigma=0.001$ which were found to work well in our experiments. Note we only present results for PSMs. Among many other supervised classifiers such as \textsc{AdaBoost} 
, \textsc{Logistic Regression} and \textsc{Support Vector Machines (Svm)}, \textsc{Random Forest (RF)} with 200 estimators and `entropy' criterion, achieved the best performance. Therefore, for brevity we only report results when \textsc{RF} is used as the classifier. 


We present results for the proposed community detection-based framework and causal and decay-based metrics. For computing $k$ nearest neighbors, we set $k=10$ as it was found to work well for our problem. By reporting the results of \textsc{Knn} trained on the decay-based causality features, we stress that using \textsc{Knn} alone does not yield a good performance. For the sake of fair comparison, all approaches were implemented and run in Python 2.7x, using the scikit-learn package. 
For any approach that requires special tuning of parameters, we conducted grid search to choose the best set of parameters.





\subsection{Baseline Methods}

\subsubsection{\textsc{Causal}~\cite{Causal2017}} We compare our metrics against the ones in~\cite{Causal2017} via supervised and community detection settings.

\subsubsection{\textsc{SentiMetrix-Dbscan}~\cite{7490315}} This was the winner of the DARPA challenge. It uses several features such as tweet syntax (e.g., average number of hashtags, average number of links), tweet semantics (e.g., LDA topics), and user behavior (e.g., tweet frequency). We perform 10-fold cross validation and use a held-out test set for evaluation. This baseline uses a seed set of 100 active and 100 inactive accounts, and then use \textsc{Dbscan} clustering algorithm to find the associated clusters. Available labels are propagated to nearby unlabeled users in each cluster based on the Euclidean distance metric, and labels of the remaining accounts are predicted using \textsc{Svm}.

\subsubsection{\textsc{SentiMetrix-RF}} This is a variant of~\cite{7490315} where we excluded the \textsc{Dbscan} part and instead trained \textsc{RF} classifier using only the above features to evaluate the feature set.

\subsection{Identification of PSM Accounts}
\begin{figure*}[t]\center
	\includegraphics[width=0.24\textwidth]{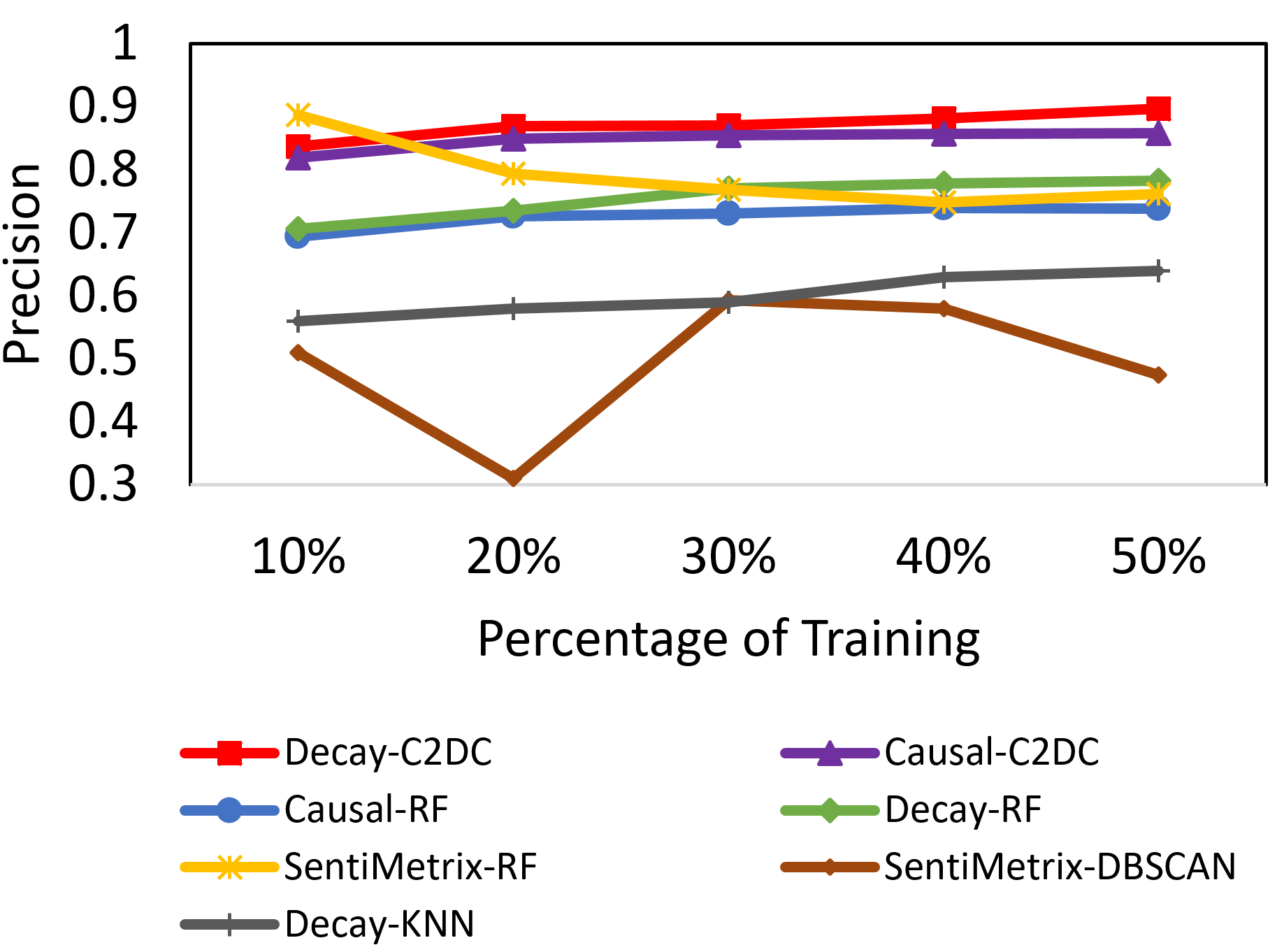}
	\includegraphics[width=0.23\textwidth]{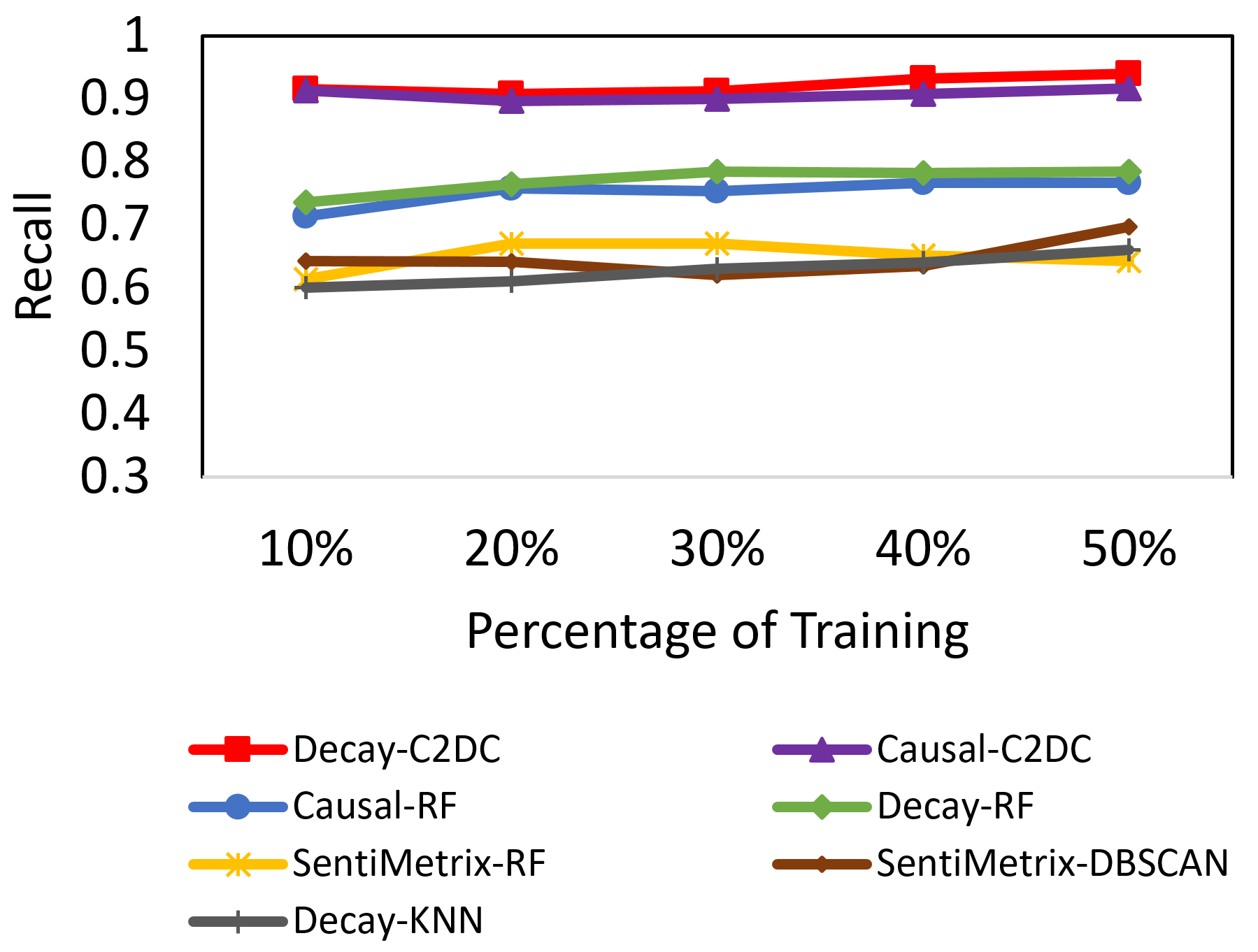}
	\includegraphics[width=0.24\textwidth]{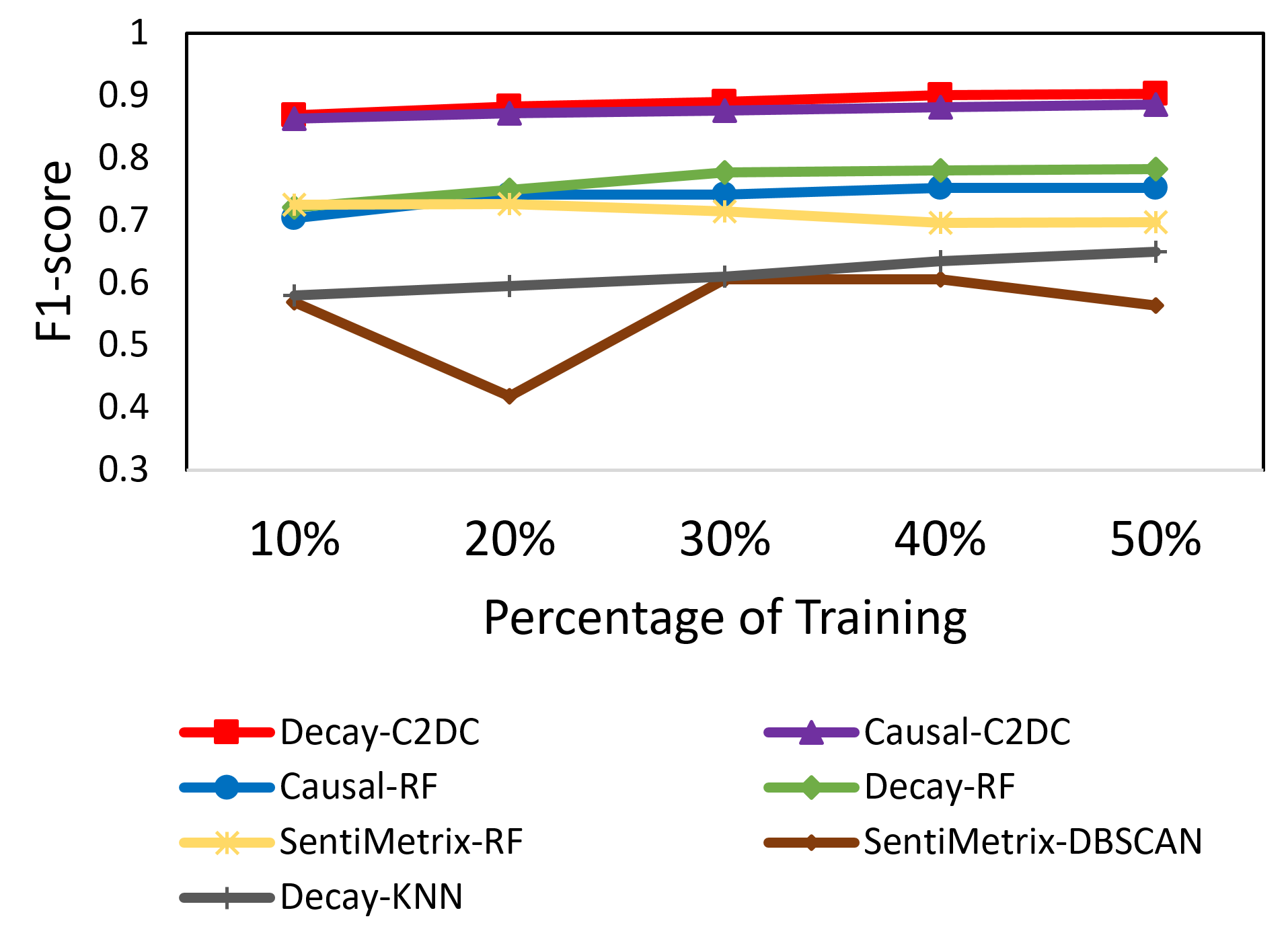}
	\includegraphics[width=0.24\textwidth]{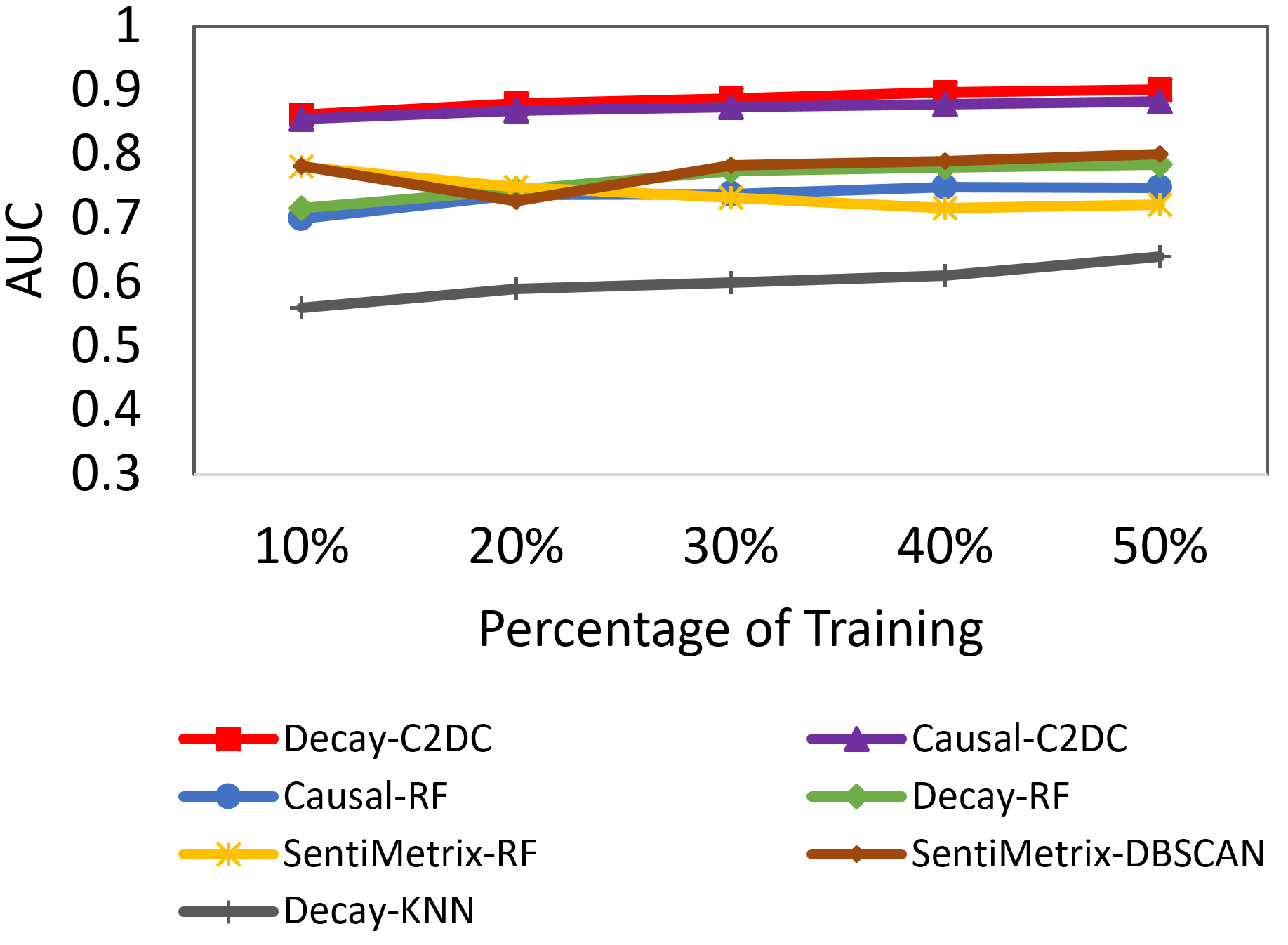}	
	\caption{Precision, recall, F1-score and AUC results for each classifier. Experiments were run using 10-fold cross validation.}
	\label{fig:results}
\end{figure*}


For each subset a separate 10-fold cross validation was performed (Figure~\ref{fig:results}). We observe the following:

\begin{itemize}
	\item Community detection achieves the best performance using several metrics. This aligns well with the $t$-test results discussed earlier: \textit{taking into account community structure of PSMs can boost the performance}.
	
	\item Causal and decay-based metrics mostly achieve higher performance than other approaches via both settings.
	
	\item Decay-based metrics are effective at identifying PSMs at different intervals via both settings. This lies at the inherent difference between decay-based and causal metrics-- our metrics take into account time-decay effect.
	
	\item Although both variants of \textsc{SentiMetrix-Dbscan} use many features, they were unable to defeat our approach. 
	
	

	
\end{itemize}

\subsection{Timeliness of PSM Accounts Identification} 
For each approach, we would like to see \textit{how many} of PSMs who were active in the first 10 days of the dataset, are correctly classified (i.e., true positives) as time goes by. Also, we need to keep track of false positives to ensure given approach does not merely label each instance as positive-- otherwise a trivial approach that always label each instance as PSM would achieve highest performance. We are also interested to figure \textit{how many} days need to pass to find these accounts. We train each classifier using 50\% of the first portion of the dataset, and use a held-out set of the rest for evaluation. Next, we pass along the misclassified PSMs to the next portions to see how many of them are captured over time. We repeat the process until reaching 50\% of the dataset-- each time we increase the training set by adding new instances of the current portion. 

There are 14,841 users in the first subset from which 3,358 users are PSMs. Table~\ref{tb:efficiency} shows the number of users from the first portion that (1) are correctly classified as PSM (out of 3,358), (2) are incorrectly classified as PSM (out of 29,617), over time. Community detection approaches were able to detect all PSMs who were active in the first 10 days of our dataset, no later than a month from their first activity. \textsc{Decay-C$^2$dc}, identified all of these PSMs in about 20 days since the first time they posted a message. Also, both causal and decay-based metrics when fed to \textsc{RF} classifier, identified all of the PSMs in the first period. \textsc{SentiMetrix-Dbscan} and \textsc{SentiMetrix-RF} failed to detect all PSMs from the first portion, even after passing 50 days since their first activity. Furthermore, these two baselines generated much higher rates of false positives compared to the rest. The observations we make here are in line with the previous ones: \textit{the proposed community detection-based framework is more effective and efficient than the rivals}. 

\begin{table*}\small
	\centering
	\caption{True/false positives for PSM accounts. Numbers are out of 3,358/29,617 PSM/Normal accounts from the first period. Last column shows the number of PSM accounts from the first period which were not caught.}
	\begin{tabular}{|c|c|c|c|c|c|c|} 
		\cline{1-7}
		\textbf{Learner} & \multicolumn{5}{c|}{\textbf{True Positives/False Positives}} & \textbf{Remaining}\\
		\hhline{=======}
		& 02/22-03/02 & 03/02-03/12 & 03/12-03/22 & 03/22-03/31 & 03/31-04/09 & \\ \cline{1-7}
		\textsc{Decay-C$^2$dc} & 3,072/131 & 286/0 & 0/0 & 0/0 & 0/0 & 0\\ \hline
		\textsc{Causal-C$^2$dc} & 3,065/156 & 188/20 & 105/0 & 0/0 & 0/0 & 0\\ \hline
		\textsc{Decay-Knn} & 2,198/459 & 427/234 & 315/78 & 109/19 & 96/0 & 213 \\ \hline
		\textsc{Decay-RF} & 2,472/307 & 643/263 & 143/121 & 72/68 & 28/0 & 0\\ \hline
		\textsc{Causal-RF} & 2,398/441 & 619/315 & 221/169 & 89/70 & 51/0 & 0\\ \hline				
		\textsc{SentiMetrix-RF} & 2,541/443 & 154/0 & 93/0 & 25/0 & 14/0 & 531 \\ \hline
		\textsc{\textsc{SentiMetrix-Dbscan}} & 2,157/2,075 & 551/5,332 & 271/209 & 92/118 & 72/696 & 215\\ \hline
		
	\end{tabular}
	\label{tb:efficiency}
\end{table*}

\section{Related Work}


\noindent \textbf{Social Spam/Bot Detection.} Recently, DARPA organized a Twitter bot challenge to detect ``influence bots''~\cite{7490315}. The work of~\cite{Cao:2014:ULG:2660267.2660269}, used similarity to cluster accounts and uncover groups of malicious users. The work of~\cite{varol2017online} presented a supervised framework for bot detection which uses more than thousands features. 
Our work does not deploy any types user related attributes. For a comprehensive survey on the ongoing efforts to fight social bots, we direct the reader to~\cite{ferrara2016rise}.

\noindent \textbf{Fake News Identification.} A growing body of research is addressing the impact of bots in manipulating political discussion, including the 2016 U.S. presidential election~\cite{shao2017spread} and the 2017 French election~\cite{ferrara2017}. For example,~\cite{shao2017spread} analyzes tweets following recent U.S. presidential election and found evidences that bots played key roles in spreading fake news.  


\noindent \textbf{Extremism Detection.} The work of~\cite{DBLP:journals/corr/KlausenMZ16} designed a behavioral model to detect extremists. Authors in~\cite{benigni2017online} performed iterative vertex clustering and classification to identify Islamic Jihadists on Twitter. Our work also differs from these works as we do not use network/user attributes.

\noindent \textbf{Detection of Water Armies.} Works of~\cite{DBLP:journals/corr/abs-1111-4297,wang2014detection} used user behavioral and domain-specific attributes to detect water armies. 

\noindent \textbf{Causal Reasoning.} As opposed to~\cite{DBLP:journals/corr/abs-1205-2634,DBLP:journals/corr/StantonTJVCS15,Kleinberg:2011:LCI:2283516.2283556} which deal with preconditions as single atomic propositions, we use rules with preconditions of more than one atomic propositions. Also, neither of~\cite{DBLP:journals/corr/abs-1205-2634,Causal2017} have addressed early detection of PSMs.

\section{Conclusion}
We enriched the existing causal inference framework to suite the problem of early identification of PSMs. We proposed time-decay causal metrics which reached F1-score of 0.6 and via supervised learning identified 71\% of the PSMs from the first 10 days of the dataset. We proposed a causal community detection-based classification algorithm, by leveraging community structure of PSMs and their causality. We achieved the precision of 0.84 for detecting PSMs within 10 days around their activity; the misclassified accounts were then detected 10 days later. Our future plan includes exploring other community detection algorithms and other forms of causal metrics. 

\section{Acknowledgments}
This work was supported through DoD Minerva program.


\end{document}